# Void Galaxy Distribution: A Challenge for ΛCDM

Saeed Tavasoli
Physics Department, Kharazmi University, Tehran, Iran; tavasoli@ipm.ir

## Abstract

We extract void catalogs from the Sloan Digital Sky Survey Data Release 16 (SDSS DR16) survey and also from the Millennium simulation. We focus our comparison on distribution of galaxies brighter than $M_r < -18$ inside voids and study the mean separation of void galaxies, distance from the void center, and the radial density profile. We find that mean separation of void galaxies depends on void size, as bigger voids have lower mean separation in both samples. However, void galaxies in the observation sample seem to have generally larger mean–distance than simulated ones at any given void size. In addition, observed void galaxies tend to reside closer to the void center than those in the simulation. This discrepancy is also shown in the density profile of voids. Regardless of the void size, the central densities of real void profiles are higher than the ones in the predicted simulated catalog.

## 1. Introduction

The internal structure of deepest under-densities of the Cosmic Web provides an ideal testing ground for the theories of cosmological structure formation. Void environments as the largest observable structures were first reported by Jöeveer et al. (1978), Tarenghi et al. (1978), Tifft & Gregory (1978), and Tully & Fisher (1978) in the three-dimensional (3D) distribution of a spectroscopic galaxy sample. Their sizes span tens to hundreds of Mpc, and close to 60% of the volume of the universe is occupied by these huge structures (da Costa et al. 1988; Geller & Huchra 1989), yet they contain only about 7% of the observed galaxies in the Universe (Pan et al. 2012).

These under-dense regions that exist between the filaments, cluster, and walls of galaxies are powerful cosmological laboratories and have been widely used in recent years to extract cosmological information (see, e.g., Pisani et al. 2015; Hamaus et al. 2016; Planck Collaboration et al. 2016, and references therein). They are expected to be less evolved and therefore better preserve the memory of the initial Universe. For instance, because voids are in the linear regime, this makes them attractive to theoretical modeling and tests. Also, several key results such as measurements of the velocity and density profiles of cosmic voids (Paz et al. 2013; Hamaus et al. 2016), the void auto-correlation function and clustering bias (Clampitt et al. 2016), the depth, abundance, and general properties of voids (Sutter et al. 2012; Nadathur et al. 2019), and the Alcock-Paczynski test (Sutter et al. 2012; Hamaus et al. 2016; Mao et al. 2017), have been obtained from these observations and simulation samples.

On the other hand, the low frequency of galaxy mergers inside these regions makes them an ideal environment for the study of the environmental dependence of galaxy formation/evolution with the minimal impact from mergers (see van de Weygaert & Platen 2011, for a review). Observationally, void galaxies have shown different properties than galaxies in more average or denser regions of the Universe. They are expected to have lower stellar mass (Croton et al. 2005; Hoyle et al. 2005; Moorman et al. 2015), tend to be bluer and have a later morphological type (Grogin & Geller 2000; Rojas et al. 2004; Patiri et al. 2006; Park et al. 2007; von Benda-Beckmann & Müller 2008; Hoyle et al. 2012; Tavasoli et al. 2015), are more gas-rich (Kreckel et al. 2012), and have lower gas-phase metallicities than those in denser regions (Mouhcine et al. 2007; Nicholls et al. 2014; Kreckel et al. 2015; Douglass & Vogeley 2017). However, there is no significant difference between the faint end slopes in the luminosity function between void and wall galaxies (Croton et al. 2005). Observational studies demonstrate that statistically void galaxies have stellar disks with smaller radii compared to a control sample of late-type galaxies (van de Weygaert & Platen 2011). Rojas et al. (2005) indicated that void galaxies are forming stars at a high rate and are fainter compared to the ones in the cluster regions.

However, the existence of low-mass galaxies in voids seems to be in conflict with the prediction of ΛCDM models (e.g., Dekel & Silk 1986; Hoffman et al. 1992; Tikhonov & Klypin 2009; Tavasoli et al. 2013). This problem was termed the "void phenomenon" by Peebles (2001).

In this Letter we study galaxy distributions inside voids to discern whether or not void environment has an influence on the geometrical position of void galaxies. We attempt to address this issue by calculating the distance of void galaxies from center of the host void and also by finding the minimum distance of void galaxies in each void separately. To achieve this we apply the Minimal Spanning Tree (MST) algorithm on a 3D void catalog extracted from observations and simulations. The observational measurements of galaxy position inside voids can be compared to the numerical orsemi-analytic simulations to test the ΛCDM model. We study this problem by analyzing voids in the Sloan Digital Sky Survey Data Release 16 (SDSS DR16; Ahumada et al. 2020) and by carrying out a parallel and comparative analysis on the Millennium simulation (Springel et al. 2005).

The outline of this Letter is as follows. In Section 2, we briefly describe the simulation and observation sample that we use in this Letter and the method that we apply to find voids. In Section 3, we introduce two parameters to investigate the distribution of void galaxies in observation and simulation sample and compare the density profile of void galaxies in both sample. We present our conclusions in Section 4.

## 2. Sample and Catalog

We study the distribution of galaxies in under-dense environments using simulated galaxies from Millennium cosmological simulation (Springel et al. 2005), and also an observed galaxy sample drawn from SDSS DR16 (Ahumada et al. 2020). The large size of the Millennium sample allows us to evaluate the properties of large-scale structure much more statistically than from observational samples. The details of the sample selection and preparation of void catalog are presented in the following.

### 2.1. Observational Sample

To find the distribution of galaxies in observed under-dense environments, we use a catalog of voids extracted from a volume-limited spectroscopic sample drawn from SDSS DR16 (Ahumada et al. 2020) using the method described in Tavasoli et al. (2013). The boundaries of our selected region of SDSS are: $130 <$ R. A. $< 235$ and $0 <$ decl. $< 55$, which contains $\sim$40,000 galaxies up to $z \sim 0.04$. The redshift of all selected galaxies are corrected for the motion of the Local Group and are given in the cosmic microwave background (CMB) restframe. Furthermore, the k-corrections of SDSS galaxies are carried out using the correct algorithm developed by Blanton et al. (2003) and Blanton & Roweis (2007). The upper limit for the redshift is defined by the limiting magnitude $M_r \sim -18$ and leaves $\sim$30,000 galaxies in the final sample to produce a homogeneous sample of data suitable for the statistical study of void galaxies. The absolute magnitudes of the galaxies are determined in the r-band using cosmological parameters; $h = 0.70$ and the density parameters $\Omega_\Lambda = 0.73$ and $\Omega_m = 0.27$ which are consistent with the millennium simulation used in this study.

### 2.2. Millennium Sample

In order to compare void-galaxy distribution in the observation and simulation samples, we select our simulated void galaxies from the semi-analytic model of galaxy formation developed by Guo et al. (2011), which has been applied to the Millennium run simulation (Boylan-Kolchin et al. 2009). This simulation was performed with cosmological parameters adopted from the Wilkinson Microwave Anisotropy Probe (WMAP) 7 data release (Komatsu et al. 2011), that is, a flat cosmological model with a non-vanishing cosmological constant $\Lambda$CDM model with $\Omega_\Lambda = 0.728$, $\Omega_m = 0.272$, $\Omega_b = 0.045$, $h = 0.70$, n = 0.96, and $\sigma_8 = 0.807$. The simulation box is 500 Mpc h$^{-1}$ on each side using $2160^3$ particles. To make a comparable result with the observation sample, we selected all simulated galaxies brighter than $\sim -18$ in the r-band filter and that lie in the $z \sim 0$ as the observational galaxy sample. This leaves $\sim 5 \times 10^6$ galaxies in our final simulated sample.

### 2.3. Void Galaxy Catalog

To construct a void catalog from the SDSS spectroscopic and Millennium simulation samples, we apply the void finder algorithm introduced originally by Aikio & Mähönen (1998; hereafter the AM algorithm) which is updated on the 3D version in Tavasoli et al. (2013). For this algorithm, the void shape is not required to be spherical (see Colberg 2008 for a review on different void algorithms). The AM algorithm works on a Cartesian grid sample to split all cells into empty and filled cells. Then, all filled cells are distinguished between wall and field galaxies by considering their distances to the third-nearest neighbor (Hoyle & Vogeley 2002). Each empty cell is assigned to a subvoid by applying the climbing algorithm (Schmidt et al. 2001) to reach the grid cell with the locally largest distance to the wall. Two subvoids are combined into a single void when the distance between them is less than their distances to the walls. Finally, all field galaxies residing within voids are classed as void galaxies (see Tavasoli et al. 2013 for further algorithm details). The void volume is estimated using the number of grid points inside a given void multiplied by the volume associated with the grid cell. For each void, we determine the void center by the positions of the grid points that enclose an elementary cell. Following this standard method and giving the same weight to all elementary cells, the center of each void can be written as

$$X_V^j = \frac{1}{N}\sum_{i=1}^{N} x_i^j$$

where $x_i^j$ ($j = 1; 2; 3$) are the locations of elementary cells and N is the number of cells in the void V.

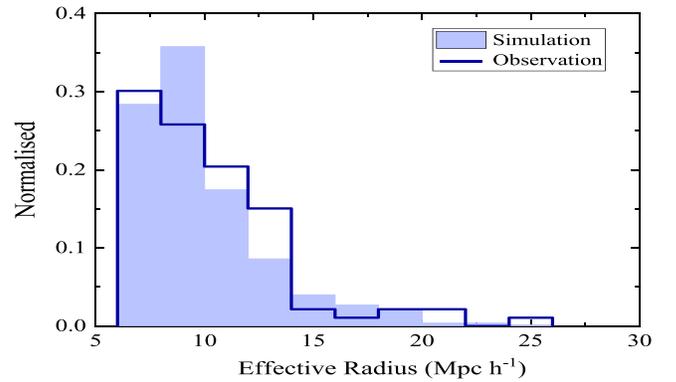

**Figure 1.** Void size distributions of observation and simulation catalogs.

The generated void catalogs in observation and simulation data, include variety of voids in size that are not in general spherical. Void sizes are characterized by an effective radius, $R_v$, corresponding to the radius of a sphere of equal volume to the void. As described in Tavasoli et al. (2013), the final void-finding algorithm excludes spurious voids with radii $R_v < 7$ Mpc h$^{-1}$ to have extremely large void regions in the local universe. By selecting all voids that lie completely inside the geometrical boundaries we avoid problems due to the boundary effects, which is a serious issue for estimating the density profile of a void (Hamaus et al. 2014) and the position of galaxies inside voids. Therefore, those voids that touch the boundaries in simulation and observation data are removed from our void catalog because of their under-estimated volumes and distorted shapes. Moreover, we consider those voids to contain at least two galaxies to estimate mean-distance void galaxies and improve our statistical analysis in this study (see Section 3.1).

This results in $\sim$100 and $\sim$18,000 voids that satisfied all of the above criteria in our observation and simulation catalogs, respectively, within which 512 and $\sim 2 \times 10^5$ void galaxies, brighter than $M_r = -18$, reside. Figure 1 presents the distribution of void size in both simulation and observation catalogs. Simulated voids are shown with a filled blue histogram and observed ones are shown with an empty dark blue histogram. It is shown that although the number of simulated voids are much more than that obtained in SDSS observations, there is a good agreement between the void size distribution of both samples.

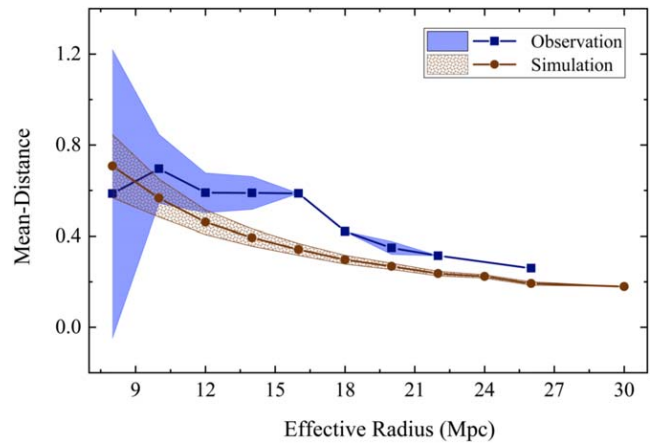

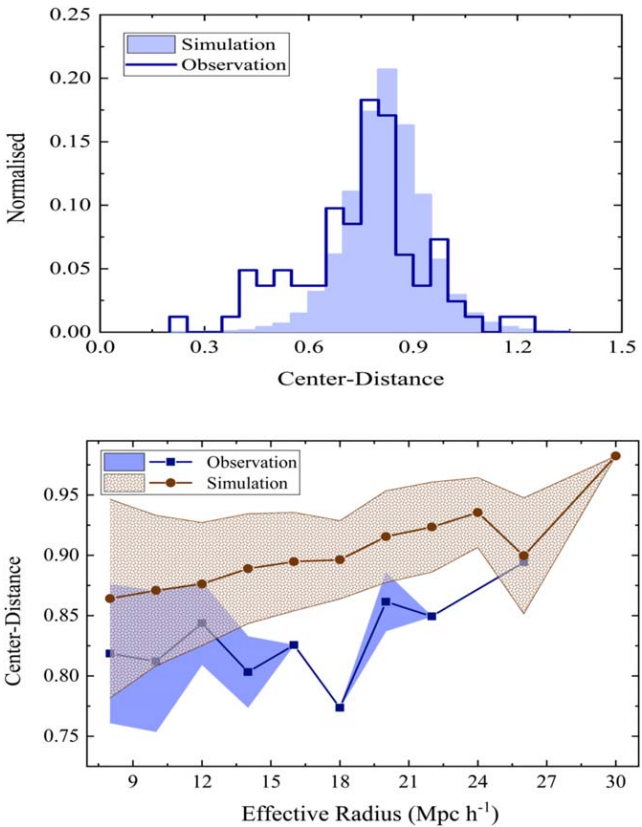

**Figure 2.** Center–distance distributions (top panel) and median center–distance as a function of void size (bottom panel) for the observation and simulation samples.

### 3. Result

#### 3.1. Galaxy Distribution

To compare the distribution of void galaxies in simulation and observation data and find possible challenges, we investigate their positions in host voids defining two specific parameters, such as center–distance and mean-distance, for both void–galaxy catalogs.

For each void the center–distance parameter is calculated by measuring the mean–distance of all void galaxies from host void center. As in our catalog there are variety of voids in each size range, which we re-scale to the $R_{\rm eff}$, where $R_{\rm eff}$ is the effective radius of the void corresponding to a spherical void with the same volume as the real object. This allows us to distinguish the position of galaxies inside all of the voids in the observation and simulation samples.

The top panel of Figure 2 illustrates the distribution of the center–distance in the observation and simulation catalogs. It shows that although most of the void galaxies are located near the boundary of voids in both samples, there is a significant difference at the center of these voids, as observed void galaxies tend to reside closer to the center of the voids than those in the simulation. Moreover, simulated void galaxies indicate generally a much narrower distribution than observed ones: the median of center–distance is ∼0.8 in observed void galaxies, whereas it is ∼0.89 in the simulation sample. We also performed a Kolmogorov–Smirnov test quantitatively that shows the probability of the two samples having similar distributions is only about 0.001, which indicates that there is clear statistical evidence for observation and simulation void galaxies being drawn from different populations.

**Figure 3.** Median mean–distance as a function of void size for observation and simulation sample.

By considering the prediction of the simulation results, we conclude that the absence of void galaxies around the void center could provide a challenge for the theory of structure formation in the standard model. Future larger observational data will certainly enable us to draw stronger conclusions.

Moreover, we compared the center–distance versus void size in observation and simulation sample in bottom panel of Figure 2. The red circle and blue square represent the median of center–distance in each effective radius bin for simulation and observation samples, respectively. The error bars are the $1\sigma$ scatter around the medians.

The result shown in this figure suggests that by increasing the void size, the distance of void galaxies from center increases in the simulation sample. This trend is not seen for the observation sample, and the center–distance are nearly constant for those voids with $R_v < 15\,{\rm Mpc}\,h^{-1}$. However, the void galaxies in the observation sample are closer to the void center than simulated voids in all size ranges.

In order to quantify how galaxies are clustered inside voids, we use a MST algorithm based on the graph theory. In an astrophysical context, the MST algorithm has been used for finding galaxy clusters and filaments (Barrow et al. 1985; Bhavsar & Ling 1988; Plionis et al. 1992), the analysis of large-scale structures (Krzevina & Saslaw 1996), and the detection of high-energy sources (Campana et al. 2008) and extended sources in EXOSAT X-ray images (De Biase et al. 1986). In this algorithm, all points of the sample are connected with the possible path-length, which contains no closed loops (see, e.g., Prim 1957). We apply this construction to a given set of galaxies inside each void to find the minimum separations. Then for each void the mean-distance parameter is defined by dividing the total branch length of the MST and number of branches. As with the center–distance parameter, we normalized mean–distance to the effective radius of a void to receive comparable parameters for all voids in both observation and simulation samples. A smaller mean–distance parameter in a void indicates more clustering galaxies. Figure 3 shows how the mean–distance parameter changes with void size in the simulation and observation catalogs. The red circle and blue square represent the median of mean–distance in each effective radius bin for simulation and observation, respectively, with 1σ scatter around the medians as the error bars. As is clear, increasing radius decreases the mean–distance in both catalogs. Also, the mean–distance of voids in the observation sample tends to be larger those in the simulation sample for the given void size.

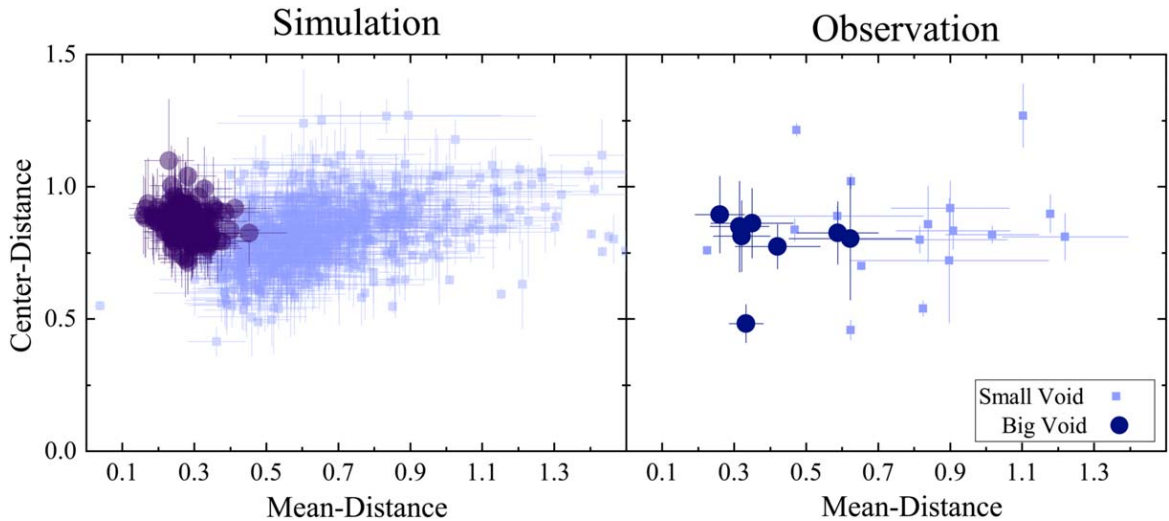

**Figure 4.** Space distribution of void galaxy (center–distance vs. mean–distance) for small (light blue) and big (dark blue) voids in simulation (left panel) and observation (right panel) sample.

Based on these results, the distribution of void galaxies predicted by ΛCDM is not consistent with the observation data. Hence, it is necessary to evaluate theories of galaxy formation in under-dense regions.

### 3.2. Galaxy Arrangement inside Different Voids

On the basis of theoretical models and computer simulations of the gravitational evolution of voids in more complex and realistic configurations, several studies have shown that as voids expand, matter is squeezed in between them, and sheets and filaments form the void boundaries. Void regions are under-densities in the mass distribution and they represent regions of weaker gravity that interpret the effective repulsive peculiar gravitational influence (Martel & Wasserman 1990; Hoffman et al. 1992; Dubinski et al. 1993; Goldberg & Vogeley 2004; Colberg et al. 2005; Padilla et al. 2005; Aragon-Calvo et al. 2010; Aragon-Calvo & Szalay 2013; Sutter et al. 2014b; Wojtak et al. 2016).

In order to understand better the dynamical evolution of voids, we studied the influence of various voids on the distribution of void galaxies and compared them with counterpart samples from simulation data. Because voids expand and increase in size, these void properties could be of potential importance for studying how galaxies populate in voids based on the ΛCDM model compared to the observation sample. To this end we selected small and large voids in both void catalogs with $R_v < 9$ Mpc h$^{-1}$ and $R_v > 15$ Mpc h$^{-1}$, respectively. This classification yielded 20 small voids (and eight large voids) in the observational sample and ∼9000 small voids (and ∼800 large voids) in the simulation sample. Figure 4 shows the distribution of void galaxies in the space of center–distance and mean–distance for the simulation (left panel) and observation (right) samples. Here, we have colored the points according to the galaxy in large and small voids as a dark and light blue, respectively. The left panel shows a clear segregation of simulated galaxies in these two specific void environments, so that most galaxies located in large voids are more concentrated than that ones in small voids, based on this space distribution.

We find that the mean value of center–distance and mean–distance of simulated large voids are ∼0.90 ± 0.06 and ∼0.32 ± 0.05, respectively, which reveals that they are living close to each other at the boundary of voids. A similar result of locating void galaxies near the void edge has been also observed by Hoyle et al. (2012) and Ricciardelli et al. (2014). This result clearly shows that in large (evolved) voids the internal structure gradually disappears as mass moves out of the void, which is in general agreement with a direct manifestation of the complex hierarchical evolution of voids by Sheth & van de Weygaert (2004) and Goldberg & Vogeley (2004). In contrast, galaxies in simulated small voids are not a homogeneous population as are the large ones and they are widely distributed in that indicated space. Specifically, they occupied most of the range of the mean–distance parameter (∼0.74 ± 0.22), from very close to very distant from void center (∼0.86 ± 0.12).

On the other hand, although we have a poor statistic in the observation sample (right panel), similar trends can be seen in the galaxy distribution of small and large voids with the simulation sample: small voids have a mean value of center–distance (∼0.86 ± 0.19) and mean–distance (∼0.94 ± 0.45), whereas those from large voids have (∼0.79 ± 0.12) and (∼0.40 ± 0.12), respectively.

### 3.3. Density Profile of Voids

The predicted form of the mean void-density profile has been proposed in several studies. Due to the different definitions of a void-finding algorithm, and also various techniques for measuring the average profile of a given set of voids, there is no unity in the functional form of the density profile. Moreover, several recent studies have come to the realization that voids in observation and simulation samples are self-similar, meaning that their average re-scaled profile does not depend on the void size and redshift (e.g., Colberg et al. 2005; Pan et al. 2012;

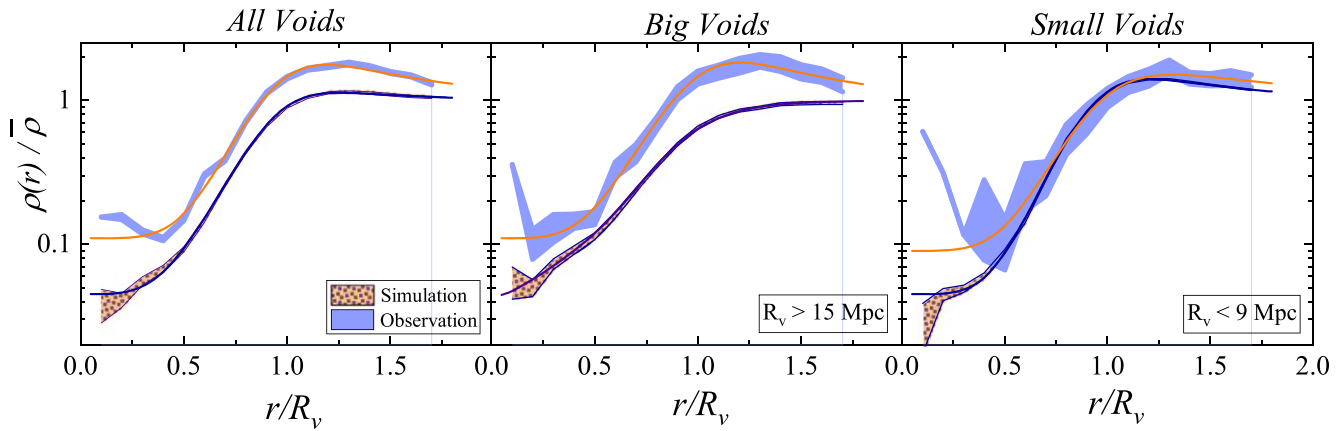

**Figure 5.** Average stacked density profiles drawn from *All-voids* (left panel), *Big-voids* (middle panel) and *Small-voids* (right panel) identified in the SDSS and simulation sample. Shaded regions depict the standard deviation $1\sigma$ within each of the stacks and the solid colored lines indicate the best fits for each curve.

Ceccarelli et al. 2013; Hamaus et al. 2014; Nadathur & Hotchkiss 2014; Ricciardelli et al. 2014; Sutter et al. 2014). By assuming self-similarity to re-scale all distances from the void center in units of the void radius, Nadathur & Hotchkiss (2014) examined the performance of three different density estimators, namely naive, VTFE, and poisson method, based on counting the number of galaxies inside voids and averaging the result over the stack of voids. They have shown that the disadvantage of "naive method" is systematically biased low in the center of small voids and the VTFE estimator works based on the voroni cell (e.g., ZOBOV algorithm) to obtain the density profile. As the generated void catalog in this study is constructed based on the AM algorithm, in the following we utilized the poisson method to investigate the density profile of voids in simulation and observation data. In order to avoid underestimating uncertainty of the density profile at distances $r \gtrsim R_v$ for voids close to the edge sample, we exclude them from both of our void catalogs (see sec Section 2.3).

Based on the poisson method, the average density in the $j$th radial bin is defined by

$$\bar{\rho}^j = \frac{(\sum_{i=1}^{N_V} N_i^j) + 1}{\sum_{i=1}^{N_V} V_i^j}$$

where $N_i^j$ is the total number of galaxies within shell volume $V_i^j$ at scaled radial distance $r_j/R_v^i$ from the void center and the sums run over all $N_v$ voids in the stack. In order to correct the systematic bias the extra $+1$ is added (Nadathur & Hotchkiss 2014).

With the aim of assessing the dependence of the void-density profiles on void radius, we classified voids into three different categories in the real and simulated void catalogs: these are *All-void* (full sample), and two subsamples split according to the void size, *Big-void* ($R_v > 15$ Mpc h$^{-1}$) and *Small-void* ($R_v < 9$ Mpc h$^{-1}$).

The result of the comparison between simulated and observed stacked void profiles is shown in Figure 5 for three defined samples, where shaded regions depict the standard deviation $1\sigma$ among all $N_v$ voids within each stack. As expected in all defined samples, stacked voids are deeply under-dense inside and the number of galaxies increases when increasing the distance from the center to the wall. This figure clearly shows that there is very good agreement between the simulated and observed stacked profiles of small voids with the expectation near the void center, $r < 0.5 R_V$. On the other hand, it is shown that the mean density of simulated voids for the *Big-void* and *All-void* samples are lower than the observed ones at all distances from the center. More interestingly, the void-density profile of the SDSS sample reveals a greater abundance of objects at the center than simulated ones without any dependence on the radius. Although there is a poor statistic to construct the stacked profile of observed *Small-void* and *Big-void* sample, which could effect the noisy shape of profile at the void center, we also find a further abundance of galaxies in observed voids rather than simulated ones at the stacked profile of the *All-void* sample. Hence, we cannot ascribe the behavior of profile at the void center to the inadequate sample effect.

Although there is little theoretical advance into predictions for the shapes of profiles, recently some work have been done to reproduce the profile shape using an empirical formula (e.g., Hamaus et al. 2014; Nadathur & Hotchkiss 2014; Ricciardelli et al. 2014). In this study we use a simple empirical formula described by Hamaus et al. (2014), which includes a functional form spanning both the interior void slope and the compensation region:

$$\frac{\rho_v(r)}{\bar{\rho}} = 1 + \delta_c \frac{1 - (r/r_s)^\alpha}{1 + (r/R_{\text{eff}})^\beta}$$

where $\delta_c$ is the central density contrast, $r_s$ a scale radius at which the profile density is equal to the average density, $R_{\text{eff}}$ is the scale distance of the void's compensation wall from the center, and $\alpha$ and $\beta$ describe the inner and outer slopes of the void profile.

To compare our stacked profiles of observed voids with those ones from simulations sample we employ the referred formula on three defined samples to extract fitting parameters. The best-fit values obtained via the above equation, including $1\sigma$ on the selected void subsamples are given in Table 1 and also represent with solid line in Figure 5.

One can see that all stacked voids, regardless of void size, produce very similar radial profiles. In addition, the external void profile shows significant differences between data and simulation, as the height of the compensation wall of simulated voids is generally dependent on the size of the voids. As we find that smaller voids are typically found in large overdense regions (the void-in-cloud process) causing the inner profile slope to become shallower and the wall to widen, in contrast large voids are generally surrounding in under-dense regions (the void-in-void process). This evolution of voids was first discussed in Sheth & van de Weygaert (2004). On the other hand, observed voids do not show such a trend on the stacked profile, as the height of the compensation wall and steepens $\alpha$ are generally constant at all void sizes.

**Table 1**
Best-fit Values and 1σ Uncertainties on the Parameters of Density Profile Formula For All, Big, and Small Voids From the Simulation and Observation Void Catalog

|            | All             | Big             | Small           |
|------------|-----------------|-----------------|-----------------|
|            |                 | Simulated       |                 |
| $\delta_c$ | $-0.95 \pm 0.01$ | $-0.97 \pm 0.04$ | $-0.95 \pm 0.01$ |
| $\alpha$   | $4.05 \pm 0.11$ | $0.87 \pm 0.21$ | $4.93 \pm 0.09$ |
| $\beta$    | $8.88 \pm 0.37$ | $6.14 \pm 2.02$ | $8.68 \pm 0.09$ |
| $r_s$      | $1.05 \pm 0.01$ | $9.67 \pm 1.59$ | $0.95 \pm 0.01$ |
| $R_{\rm eff}$ | $1.01 \pm 0.01$ | $0.95 \pm 0.19$ | $1.02 \pm 0.01$ |
|            |                 | Observed        |                 |
| $\delta_c$ | $-0.89 \pm 0.01$ | $-0.89 \pm 0.01$ | $-0.91 \pm 0.07$ |
| $\alpha$   | $5.01 \pm 0.11$ | $4.50 \pm 0.54$ | $4.67 \pm 1.88$ |
| $\beta$    | $8.42 \pm 0.11$ | $8.50 \pm 0.80$ | $7.43 \pm 1.15$ |
| $r_s$      | $0.88 \pm 0.01$ | $0.88 \pm 0.01$ | $0.97 \pm 0.03$ |
| $R_{\rm eff}$ | $1.04 \pm 0.01$ | $1.09 \pm 0.06$ | $1.07 \pm 0.14$ |

The interesting result of this table arises by comparing the central density contrast, $\delta_c$, of both void catalog. We find that the $\delta_c$ of observed voids in all subsamples are higher than that ones in the simulated catalog, which gives us information of the depletion of predicted halos by ΛCDM at the center of voids. Further detailed study by using deeper and more observation data will clarify this issue.

Many authors have discussed and presented measured radial density profiles in data and simulations, and there appears to be a generally universal shape to the profile (e.g., Benson et al. 2003; Padilla et al. 2005; Ceccarelli et al. 2006; Lavaux & Wandelt 2012; Sutter et al. 2012).

However, there is a difference between the value of fitting parameters obtained here with those in other studies. This is due to the use of different void-finding algorithms, techniques for measuring the average profile of a given set of voids, and also sample selection.

## 4. Summary and Conclusion

We have performed a statistical study of the distribution of galaxies focusing on void environments in the local universe. To this aim, we have carried out a parallel study of the voids in the SDSS DR16 redshift survey and the semi-analytic model of galaxy formation based on the Millennium run simulation. As a matter of comparison, we apply a similar void-finding algorithm (AM) and consider the same luminosity threshold for both samples (e.g., $M_r = -18$), to construct a void catalog. In order to assess how galaxies are distributed in void environments, we utilize an MST algorithm to estimate the mean separation of galaxies inside each void and evaluate the mean–distance of galaxies from the void center.

We find that there is discrepancy between the distribution of real and simulated void galaxies close to the center, as observed void galaxies tend statistically to reside closer to the center of voids than to the prediction of a simulation sample. Furthermore, there is a correlation between mean–distance and void size in both samples, as void galaxies in larger voids have lower mean–distance. However, void galaxies in the observation sample seem to have a larger mean–distance than simulated ones at any given void size. In addition, galaxies inside large voids are located together near the boundary in both samples, whereas this configuration is not seen obviously in small voids. We have examined the universality of void-density profiles of real and simulated voids. We found that all profiles in both samples can be well described by the functional form proposed in Hamaus et al. (2014). It is shown that the compensation wall of simulated voids is generally dependent on void size as this trend does not appear for observed ones. Moreover, the inner and outer slopes of the void profiles, regardless of the void size, are different in real and simulated void catalogs and also the central density of the real void profile is higher than that ones in the predicted simulated catalog.

Therefore, we argue that the difference at the center of the density profile between an observed and simulated void is a robust result and is due to the depletion of galaxies closer to the simulated void center. This result provides a reference point for comparisons with theoretical models and may prove to be a useful observational tool to constrain cosmology. For instance, Novosyadlyj et al. (2017) found that although energy perturbation negligibly affects the universal density profiles, the different initial conditions result in voids with variety central density and also the existence of void galaxies at different distance scales from the void center. This means that the investigation of the internal structure of voids could provide a useful perspective for testing models of dark energy and gravity modifications.

To corroborate this hypothesis, we need more observed voids, and other hydrodynamic cosmological simulations (such as the Illustris; Vogelsberger et al. 2014), to see if the marginal difference reported here is of any significance. Our work suggests that the internal structures of voids provide a powerful framework to constrain models of cosmology that can be exploited in future studies of voids with upcoming large surveys such as LSST and Euclid.

Although this is an important topic to study, both observationally and theoretically, more explorations are needed as far as defining and finding voids to investigate whether or not there is a significant difference between empirical and simulated void distributions dependent on the definition of a void algorithm.

This Letter has made use of SDSS data. Funding for the SDSS and SDSS-II has been provided by the Alfred P. Sloan Foundation, the Participating Institutions, the National Science Foundation, the U.S. Department of Energy, the National Aeronautics and Space Administration, the Japanese Monbukagakusho, the Max Planck Society, and the Higher Education Funding Council for England. The SDSS Web site is http://www.sdss.org/. The SDSS is managed by the Astrophysical Research Consortium for the Participating Institutions. The Participating Institutions are the American Museum of Natural History, Astrophysical Institute Potsdam, University of Basel, University of Cambridge, Case Western Reserve University, University of Chicago, Drexel University, Fermilab, the Institute for Advanced Study, the Japan Participation Group, Johns Hopkins University, the Joint Institute for Nuclear Astrophysics, the Kavli Institute for Particle Astrophysics and Cosmology, the Korean Scientist Group, the Chinese Academy of Sciences (LAMOST), Los Alamos National Laboratory, the Max-Planck-Institute for Astronomy (MPIA), the MaxPlanck-